\documentclass[conference]{IEEEtran}

\makeatletter
\def\ps@headings{%
\def\@oddhead{\mbox{}\scriptsize\rightmark \hfil \thepage}%
\def\@evenhead{\scriptsize\thepage \hfil \leftmark\mbox{}}%
\def\@oddfoot{}%
\def\@evenfoot{}}
\makeatother

\usepackage{cite}
\usepackage{amsmath}
\usepackage{amssymb}
\usepackage{xspace}
\usepackage{graphicx}
\usepackage{url}
\usepackage[algoruled,vlined]{algorithm2e} 

\graphicspath{eps/}

\SetAlgoSkip{bigskip}

\renewcommand{\algorithmcfname}{Strategy}%

\renewcommand{\subsubsection}[1]{\smallskip\noindent{\bf #1}}

\newcommand{\ie}{{\em i.e.}\xspace}
\newcommand{\flickr}{{\em Flickr}\xspace}
\newcommand{\contact}{{\em contact}\xspace}
\newcommand{\symmcontact}{{\em symmetric-contact}\xspace}
\newcommand{\comment}{{\em comment}\xspace}
\newcommand{\symmcomment}{{\em symmetric-comment}\xspace}

\newcommand{\ran}{\ensuremath{\mbox{ran}}}

\newcommand{\degreeprinciple}{{\em degree principle}\xspace}
\newcommand{\triangleprinciple}{{\em triangle principle}\xspace}

\def\set#1{\ensuremath{\{ #1 \}}}

\def\mathwr#1#2{\ensuremath{\mathtt{#1}(#2)}}

\def\V{\ensuremath{\vee}}
\def\tri{\ensuremath{\Delta}}
\def\gcc#1{\mathwr{tr}{#1}}
\def\lcc#1{\mathwr{cc}{#1}}

\def\strat#1#2{\ensuremath{\textit{#1}_{#2}}}

\def\eff{\ensuremath{\mathcal{E}}}
\def\reff{\ensuremath{\mathcal{R}}}

\def\reff{\ensuremath{\mathcal{R}}}

\begin{document}

\title{Efficient Measurement of Complex Networks\\ \smallskip Using Link Queries}

\author{\IEEEauthorblockN{Fabien Tarissan\,\thanks{Contact author -- \url{tarissan@lix.polytechnique.fr}}}
\IEEEauthorblockA{ISC\\ CNRS and \'Ecole Polytechnique\\ \url{tarissan@lix.polytechnique.fr}}
\and
\IEEEauthorblockN{Matthieu Latapy}
\IEEEauthorblockA{LIP6\\CNRS and Universit\'e Pierre et Marie Curie\\ \url{Matthieu.Latapy@lip6.fr}}
\and
\IEEEauthorblockN{Christophe Prieur}
\IEEEauthorblockA{LIAFA\\Universit\'e Paris Diderot\\ \url{prieur@liafa.jussieu.fr}}
}

\maketitle

\begin{abstract}
Complex networks are at the core of an intense research activity. However, in most cases, intricate and costly measurement procedures are needed to explore their structure. In some cases, these measurements rely on link queries: given two nodes, it is possible to test the existence of a link between them. These tests may be costly, and thus minimizing their number while maximizing the number of discovered links is a key issue. 
This paper studies this problem: we observe that properties classically observed on real-world complex networks give hints for their efficient measurement; we derive simple principles and several measurement strategies based on this, and experimentally evaluate their efficiency on real-world cases. In order to do so, we introduce methods to evaluate the efficiency of strategies. We also explore the bias that different measurement strategies may induce.

\end{abstract}




\section{Preliminaries}
\label{sec-intro}

Complex networks, modeled as large graphs, are everywhere in science, society, and everyday life. 
However, it must be clear that most real-world complex networks are not directly available: collecting information on their structure generally relies on intricate and expensive measurement procedures. Conducting such a measurement often is a challenge in itself, and is an important part of the work needed to study a complex network.

In general, complex network measurements consist in a combination of a few simple measurement primitives. In several cases, this primitive consists in testing the existence of a link, which we call a {\em link query}: given two nodes $u$ and $v$, a measurement operation makes it possible to decide whether there is a link between them or not. This simple test may be expensive (regarding the needed resources or time, or the load it induces on the network, for instance) and so conducting measurements with as few calls to the measurement primitive as possible is a key issue.

For instance, in online social networks like Facebook or Flickr\,\footnote{\url{http://www.facebook.com/} and \url{http://www.flickr.com/}}, privacy concerns and reduction of server load often lead to limitations in the queries that one is allowed to perform to explore networks between users. Link queries are however allowed in most cases. Likewise, measurements of real-world social networks often rely on interviews, in which link queries play a central role\,\cite{wellman07personal}. In biological networks like protein interactions or gene regulatory networks, link queries also play a key role \cite{kucherov05bioqueries, kepes02topo}.

In all these contexts, and others, link queries are very expensive: they have a significant load on server running online social network software and their number is generally bounded; they have a significant cost for interviewers and participants in sociological studies ; or they require costly biological experiments, depending on the case.

\medskip

In this paper, we formalise this problem as follows: given a graph $G=(V,E)$, we want to define {\em strategies} (ordered lists of link queries) which lead to the discovery of as many links of the network as possible. In other words, we want to minimize the number of link queries while maximizing the number of observed links, i.e. the number of positive answers to these tests\,\footnote{Notice that, whereas we suppose that link queries are very expensive, the computational cost of each strategy is not our concern here; we consider it as negligible compared to measurement costs, which fits most real-world cases.}.

In order to do so, we will rely on simple intuitions derived from statistical properties observed on most real-world complex networks, which we discuss in Section~\ref{sec-principles}. We then propose several measurement strategies in Section~\ref{sec-strategies} based on these principles. We also need a way to compare and evaluate measurement strategies, see Section~\ref{sec-evaluation}. We finally use this to experimentally evaluate proposed strategies in Section~\ref{sec-experiments}.

Before entering in the core of this paper, we give the needed formalism and notations, and discuss related work.

\subsection{Formalism and notations}

In all the paper, we will consider an undirected\,\footnote{This means that we make no difference between $(u,v)$ and $(v,u)$, for any $u$ and $v$.} graph $G = (V,E)$, with $n = |V|$ nodes and $m=|E|$ links. We suppose that all the nodes are known, and focus on link discovery only. In other words, we know $V$ but know nothing about $E$ (although we will make some statistical assumptions in accordance with classical empirical observations in the field, see Section~\ref{sec-principles}).

We will denote by $N(v)$ the set of neighbors of $v \in V$: $N(v) = \set{u \in V,\ (u,v) \in E}$ and by $d(v)$ its degree: $d(v) = |N(v)|$.

A measurement consists in a series of link queries, \ie\ tests of the existence of link $(u,v)$ for two nodes $u$ and $v$ in $V$. At a given stage in such a measurement, one has already discovered a set of links, which we will denote by $E' \subseteq E$. The set of extremities of links in $E'$ will be denoted by $V' \subset V$. Notice that, although we know $V$, in general $V' \not=V$. We will also denote by $n'$ the number of nodes in $V'$ and $m'$ the number of discovered links so far: $n' = |V'|$ and $m' = |E'|$. We also define $N'(v) = N(v)\cap V'$ and $d'(v)=|N'(v)|$ for all $v\in V'$. Notice that both $V'$, $E'$, $n'$, $m'$, $N'$ and $d'$ vary during a measurement; however, the context will make it clear which value we consider.

\subsection{Related work}

This work belongs to the fields of complex network metrology, which mostly focused on the specific case of the Internet topology until now, see for instance \cite{lakhina02sampling,achlioptas05bias,guillaume2005relevance,latapy08complex,stutzbach06sampling,stutzbach06unbiased,dallasta04statistical}. This area of research aims mainly at evaluating the relevance of collected complex network samples and properties observed on them, and correcting these observations. Viewing the measurement as the combination of many instance of a simple primitive (link queries, here) which we want to optimize is new, and is an important contribution of this paper.

Another related problem is the one of {\em link prediction}: given a network in which new links may appear, one wants to predict which new links will appear in the future based on currently existing ones \cite{newman08hierpred, liben-nowell07linkpred}. In this context, authors use properties of the known network to infer probable future link, which is similar to what we do below in the measurement context. The main difference lies in the fact that very little of the network topology is known in our case.
\section{Underlying principles}
\label{sec-principles}

Our goal is to design measurement strategies based on link queries (test of the existence of a link between two given nodes) which will minimize the number of such queries and maximize the number of discovered links (\ie the number of positive answers to these tests). In order to do so, we will rely on some simple statistical properties which are observed on most real-world complex networks \cite{watts1998smallworld}.

\subsection{Properties of complex networks}

First, we will suppose that $G$ is sparse: its density $\delta = \frac{2.m}{n.(n-1)}$ is very small. In other words, the probability that a link exists between two randomly chosen nodes is very small, \ie a random link query will fail with high probability.

The second key property is the fact that most complex networks have a very heterogeneous degree distribution (often close to a power law). Since the degree of a node is the number of links attached to it, this means that there is a high variability between the number of links of each node (many nodes have very few links, but some have more, and even many more).

Finally, another key property is the local density: although randomly chosen nodes have a very low probability to be linked, two nodes which have a neighbor in common are linked with a much higher probability. This is generally captured by the clustering coefficient or the transitivity ratio \cite{watts1998smallworld,schank05findingWEA,schank04approximating}, defined by:
$$
\lcc{G}= \frac{\sum_v \frac{\tri(v)}{\V(v)}}{n}
$$
$$
\gcc{G}= \frac{3.\tri(G)}{\V(G)}
$$
where, for each $v\in V$, $\tri(v)$ denotes the number of triangles (sets of three nodes with three links) to which $v$ belongs; $\V(v) = \frac{d(v).(d(v)-1)}{2}$ denotes the number of pairs of neighbors of $v$; $\tri(G) = \sum_v \tri(v)$; and $\V(G) = \sum_v \V(v)$.

A classical observation in complex network studies is that both these quantities are high, at least compared to the density. In other words, if one chooses a random pair of links with an extremity in common (transitivity ratio) or a random node and two of its neighbors (clustering coefficient) then the probability that the third possible link exists is high.

\subsection{Consequences on measurements}

The properties above, observed on most real-world complex networks, have a strong impact on measurements and will play a key role here.

First, the low density of complex network implies that randomly choosing two nodes and testing the presence of a link between them is very inefficient. Notice however that, when only link queries are possible, one has no choice but to begin with a series of such random measurements. However, it must be clear that exploring a large complex network with such a strategy only is not reasonable.

Instead, the existence of nodes with degree much larger than the average may be useful for efficient measurement. Suppose that we test a random pair $(u,v)$. The probability that it is positive (\ie the link $(u,v)$ exists) is proportional to the degree of $u$ (resp. $v$). Therefore, if it exists then one may guess that $u$ (resp. $v$) has a high degree, and so testing all pairs $(u,w)$ (resp. $(v,w)$) for any $w$ will probably lead to the discovery of many links. Notice that $u$ and $v$ play a symmetric role in this reasoning. We will call this observation the \degreeprinciple.

Likewise, the high local density may be used for efficient measurement: when we know that two nodes $u$ and $v$ have a neighbor $w$ in common then testing pair $(u,v)$ certainly makes sense as this link exists with high probability. We call this the \triangleprinciple.

We may now turn to the definition of measurement strategies based on these principles.

\section{Measurement strategies}
\label{sec:meas}
\label{sec-strategies}

First notice that when one starts a measurement in our framework, no link is known and we have no way to distinguish between vertices. Therefore, there is no choice but to test random pairs of nodes. We call this null strategy \strat{random}{k}.

\begin{algorithm}[H]
\caption{\strat{random}{k} with $k$ an integer.}
\label{strat:random}
\dontprintsemicolon
\While{$m' < k$}{
 test a random untested pair\;
 }
\end{algorithm}

As soon as some links are discovered, though, one may try to design more efficient strategies. The \triangleprinciple indicates that, when a \V\ pattern is discovered one may test the missing link in the triangle. This leads to the following strategy.

\begin{algorithm}[H]
\caption{\strat{\V-random}{k} with $k$ an integer.}
\label{strat:vrandom}
\dontprintsemicolon
\While{$m' < k$}{
 Test a random untested pair $(u,v)$\;
 \If{$(u,v)$ exists}{
  Test all untested pairs $(v,w)$, for any $w$ in $N'(u)$\;
  Test all untested pairs $(u,w)$, for any $w$ in $N'(v)$\;
  }
 }
\end{algorithm}

Applying directly the \degreeprinciple would lead to a strategy in which we test the pairs $(u,v)$ for all $v$ as soon as a random test led to the discovery of a link of $u$. However, the \degreeprinciple becomes stronger if one waits until {\em several} links of a node are found. We therefore propose a strategy in which a series of tests (performed according to another strategy) is followed by a use of the \degreeprinciple on nodes for which we discovered many links.

\begin{algorithm}[H]
\caption{(\V-)Complete Simple --- \strat{cs}{k} (resp. \strat{\V-cs}{k}) with $k$ an integer.}
\label{strat:complete}
\dontprintsemicolon
Apply \strat{random}{k} (resp. \strat{\V-random}{k})\;
\ForEach{$u \in V'$ in decreasing order of $d'(u)$}{
 Test all untested pairs $(u,v)$, for any $v\in V$\;
 }
\end{algorithm}

This strategy may be improved by using the links it discovers for
choosing the next link queries to perform. This leads to the following
strategy.

\begin{algorithm}[H]
\caption{(\V-)Complete --- \strat{c}{k} (resp. \strat{\V-c}{k}) with $k$ an integer.}
\label{strat:icomplete}
\dontprintsemicolon
Apply \strat{random}{k} (resp. \strat{\V-random}{k})\;
Let $X = V'$\;
\While{$X$ is nonempty}{
 Let $u$ in $X$ with $d'(u)$ maximal\;
 Remove $u$ from $X$\;
 Test all untested pairs $(u,v)$, for any $v\in V$\;
 \If{$(u,v)$ exists and is the first link of $v$ discovered}{
  Add $v$ to $X$\;
  }
 }
\end{algorithm}

One may try to use an even stronger version of the \degreeprinciple by noticing that the probability of a link between two nodes is even larger if {\em both} have a high degree. Therefore, link queries between nodes for which we already discovered many links have an even higher probability of positive outcome. This leads to the following strategy.

\begin{algorithm}[H]
\caption{(\V-)Test-Between-Found --- \strat{tbf}{k} (resp. \strat{\V-tbf}{k}) with $k$ an integer.}
\label{strat:tbf}
\dontprintsemicolon
Apply \strat{random}{k} (resp. \strat{\V-random}{k})\;
\ForEach{$(u,v) \in V'\times V'$ in decreasing order of $d'(u)+d'(v)$}{
 Test $(u,v)$ if it was untested\;
 }
\end{algorithm}

Finally, one may try to combine the strategies above in order to improve their efficiency. Indeed, some of them use complementary principles which both help in discovering more links with less link queries. One may therefore expect even better results with combinations of them. We will therefore consider the following strategy.

\begin{algorithm}[H]
\caption{(\V-)TBF-Complete --- \strat{tbfc}{k} (resp. \strat{\V-tbfc}{k}) with $k$ an integer.}
\label{strat:tbfic}
\dontprintsemicolon
Apply \strat{tbf}{k} (resp. \strat{\V-tbf}{k})\;
Apply \strat{c}{0}\;
\end{algorithm}

It must be clear that many variants and improvements of the strategies above are possible. Probably, completely different strategies may also be defined. Our goal here however is to evaluate the relevance of the \degreeprinciple and \triangleprinciple in the design of measurement strategies. We therefore focus on these relatively simple strategies, which we consider as a natural first set of strategies derived from these basic principles.

\section{Evaluation methodology}
\label{sec-evaluation}

For any measurement strategy $S$, let us define $m_S'(q)$ as the expected number of links discovered with $q$ link queries with strategy $S$\,\footnote{Notice that, in practice, it is in general impossible to reach a situation where we test all pairs of nodes: $q = \frac{n (n-1)}{2}$, or conversely where we discovered all existing links: $m'_S(q) = m$.}. It must be clear that our goal, for a given $q$, is to design a strategy $S$ that maximises $m'_S(q)$. Conversely, one may want to discover a given number $x$ of links and ask for the strategy $S$ that will minimize the $q$ such that $m'_S(q) = x$.

However, given two numbers of queries $q$ and $r$ it is possible that a given strategy $S$ discovers more links with $q$ tests than another strategy $T$, while $T$ discovers more with $r$ tests (we will observe such a situation in Section~\ref{sec-typical}). As a consequence, it makes no sense to say that $S$ is better than $T$, nor the converse; this depends on the allowed number of link queries.

Going further, one may notice that if $S$ and $T$ discover the same number of links after a given number $q$ of tests, but if $S$ discovers more links than $T$ for any number $r<q$ of test, then it seems natural to consider that $S$ surpasses $T$ (it discovers the same number of links, but faster).

A simple way to formalise these intuitions is to define the {\em efficiency} of a strategy $S$ for a given number of queries $q$ as the (discrete) integral of the function $m'_S$ from $0$ to $q$: $\eff_q(S) = \sum_{i=1}^{q}m'_S(i)$.

\medskip

Notice that the obtained value will depend on the considered graph, and on $q$. It seems difficult to avoid this, as the efficiency of strategies do indeed depend on the graph under concern, and on the number of allowed link queries. We will therefore always compare strategies ran on the same graph and with the same number of link queries here.

Another weakness of this definition is that it may give any positive value for the efficiency of a strategy, making it hard to evaluate how far from the worst or best solution we are. In order to avoid this we introduce the {\em normalised efficiency}: $\overline{\eff}_q(S) = \frac{\eff_q(S) - \eff_q(\min)}{\eff_q(max) - \eff_q(\min)}$ where $\min$ and $\max$ stand for the worst and best strategies, \ie the ones with minimal and maximal efficiencies.

Notice that strategies $\min$ and $\max$ are easy to determine: $\min$ consists in testing pairs of nodes with no links between them as long as possible, thus $\frac{n (n-1)}{2} - m$ times, and then performing the positive tests; conversely $\max$ consists in performing first the $m$ positive tests. As a consequence, we can compute easily $\eff_q(\min)$ and $\eff_q(max)$ for any $q$, and thus obtain the normalized efficiency of any strategy.

\medskip

The notion of normalized efficiency however remains insufficient. Indeed, as we consider sparse graphs, there are only very few positive link queries, and thus one may expect to be much closer to the $\min$ strategy than to the $\max$. As a consequence, the efficiency of any strategy will be very low.

A solution to this problem consists in comparing strategies to the random one, denoted by $\ran$, which consists in performing link queries on random untested pairs of nodes. The expected efficiency of this strategy is easy to compute, as the probability of success of a link query is exactly the density $\delta$; we obtain: $\eff_q(\ran) = \sum_{i=1}^q i\ \delta = \frac{q (q+1)}{2}\ \delta$.

Finally, we introduce the {\em relative efficiency}, which indicates how a given strategy $S$ performs compared to the random one (and the minimal and maximal ones) after $q$ link queries: $\reff_q(S) = \frac{\overline\eff_q(S)}{\overline\eff_q(\ran)}$.

Notice that the relative efficiency does not give a value between $0$ and $1$ and therefore does not have the advantage of being relatively independent from the context. However, we cannot normalize it as we would lose the benefit of the comparison to the random strategy. We will therefore use both the normalized efficiency and the relative efficiency to discuss efficiency of strategies below, and keep in mind that in any case the efficiency of a strategy depends on the graph under concern and on the number of link queries allowed. Only the full $m'_S()$ function can describe the efficiency of strategy $S$ entirely, on a given graph.

\section{Experimental evaluation}
\label{sec:app}
\label{sec-experiments}

In this section, we present experiments aimed at illustrating the differences between the proposed measurement strategies, and how they may be evaluated. We first present the dataset we used, which is a typical real-world case. We then examine a typical situation and discuss the observations. We deepen this by observing the impact of the initial random period of measurement; and finally we discuss the bias that measurement strategies may induce on observed properties.

\subsection{Dataset}\label{sec:data}

We use here data on an online social network which we consider as a typical example of complex networks studied in the literature. This social network comes from the \flickr site, which provides facilities for publishing online photos, sharing them with others, discuss them, etc. Users may also subscribe to various interest groups and have lists of other users known as their {\em contacts}.

Here we used a complete measurement of \flickr\ conducted in August 2006 \cite{prieur08weakcoop}. We considered the largest of the 72\,875 groups observed then\,\footnote{{\em FlickrCentral}, \url{http://flickr.com/groups/central/}}, which contained 31\,523 members.

We then defined three different networks among these 31\,523 users:
\begin{itemize}
\item \contact: two users $a$ and $b$ are linked if $a$ is a contact of $b$ or $b$ is a contact of $a$;
\item \comment: two users $a$ and $b$ are linked if $a$ posted a
  comment on a photo from $b$ or $b$ posted a comment on a photo from
  $a$;
\item \symmcomment: two users $a$ and $b$ are linked if both $a$
  posted a comment on a photo from $b$ and $b$ posted a comment on a
  photo from $a$.
\end{itemize}

One may also define a \symmcontact graph in which two users $a$ and $b$ are linked if both $a$ is a contact of $b$ and $b$ is a contact of $a$
. In order to save space, we will not consider it here. Likewise, we do not detail the features of these networks; the key point here is that they are sparse, have heterogeneous degree distributions and high clustering coefficient and transitivity ratio. To this regard, they are similar to most real-world complex networks, and so the principles discussed in \ref{sec-principles} apply.



\subsection{A typical example}
\label{sec-typical}

\begin{figure}[h!]
\begin{center}
\includegraphics[width=8cm]{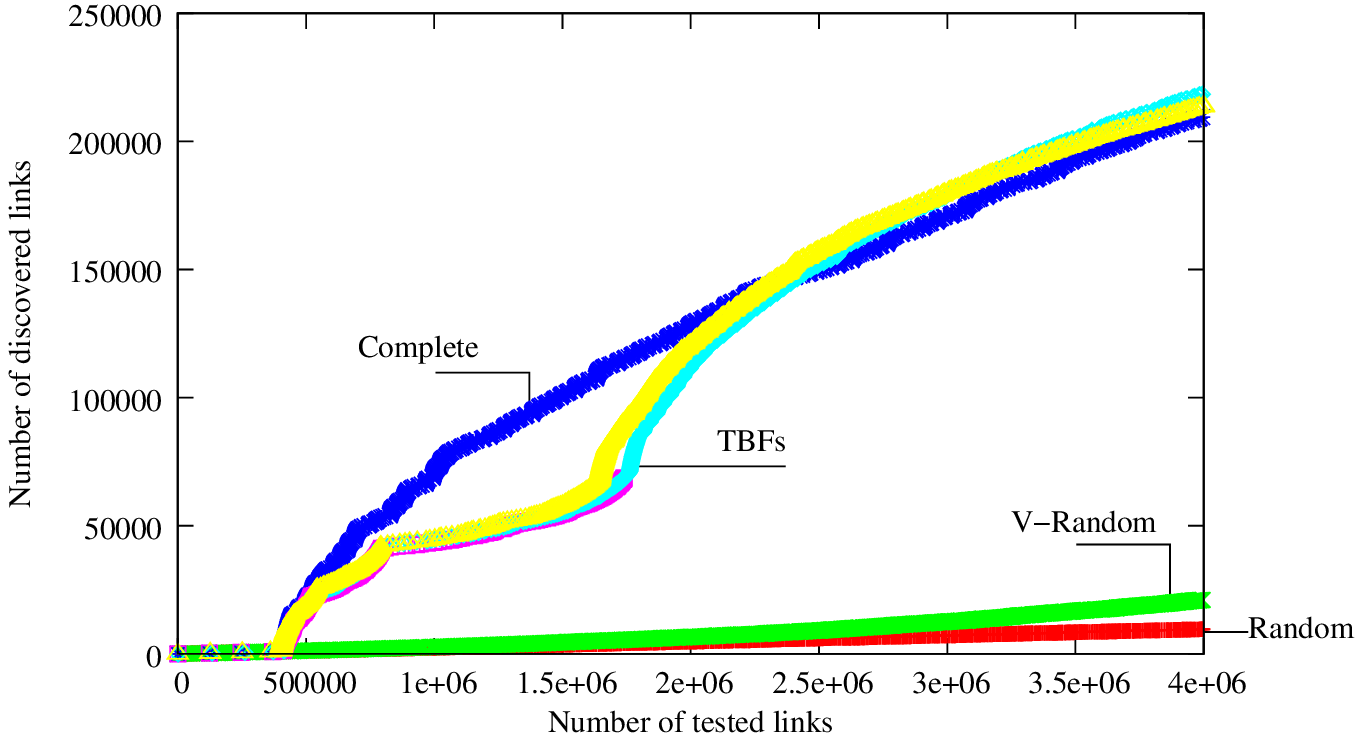}
\end{center}
\caption{
Number of links (vertical axis) discovered by each strategy as a function of the number of link queries performed (horizontal axis) in a typical case (\contact network, $k=1\,000$). The \strat{tbf}{}, \strat{tbf-complete}{} and \strat{\V-tbf-complete}{} strategies are indicated as a unique curve (named TBFs) in the plot as the three curves overlap eachother.
}
\label{fig:curve-1000}
\end{figure}

Let us first try all our strategies with the same parameter $k=1\,000$ and on the \contact graph. We represent in Figure~\ref{fig:curve-1000} the number $m'_S(q)$ of links discovered by each strategy $S$ as a function of the number $q$ of link queries performed, for $q$ between $0$ and $Q=4.10^6$. The obtained plot is representative of what is obtained on other graphs.

This plot shows clearly that measurement strategies perform very differently, and that trying to optimize them is relevant. This is a first important result in itself. Moreover, both the \degreeprinciple and the \triangleprinciple are useful in doing so: strategies based on each of them perform significantly better than the random strategy. However, the \degreeprinciple seems to be much stronger: while the improvement of \strat{$\V$-random}{} remains quite low, the improvement obtained by the \strat{complete}{} strategy is huge. This is probably due to the fact that, although the clustering coefficient and transitivity ratio are much larger than the density, they remain quite small; instead, the largest degrees in the graphs are relatively close to its number of nodes.


The best final results (the largest number of links discovered at the end of the measurement) are obtained with mixed strategies, namely \strat{tbf-complete}{} and \strat{\V-tbf-complete}{}, which succeed in discovering between $22$ and $23 \%$ of existing links by performing only $1 \%$ of possible link queries. They slightly outperform \strat{complete}{}, which was expected as they are more subtle (though a stronger improvement may have been expected).

Notice that, although these strategies finally outperform \strat{complete}{}, they discover links {\em later} than this strategy. In this sense, they may therefore be considered as less efficient, which is captured by our notion of efficiency, see Table~\ref{tab:eff-1000}.

\begin{table}[h!]
\centering
\begin{tabular}{|c|c|c|c|c|c|}
  \hline
   & $m'$ & \% tested & \% found & $\mathcal{E}$ & $\mathcal{R}$\\ 
  \hline
  \strat{random}{} & 9\,609 & 1.04 & 1.03 & 0.006 & 0.99\\
  \strat{\V-random}{} & 21\,030 & 1.04 & 2.25 &0.010 & 1.64 \\
  \strat{c}{1000} & 209\,485 & 1.04 & 22.4 & 0.142 & 24.2 \\
  \strat{tbf}{1000} & 68\,874 & 0.46 & 7.36 & 0.048 & 15.6\\
  \strat{tbfc}{1000} & 218\,448 &1.04 & 23.4 &0.131 & 22.3\\
  \strat{\V-tbfc}{1000} & 214\,175 &1.04 & 22.9 & 0.134 & 22.7\\
  \hline
\end{tabular}
\caption{\em \em
Efficiency of each strategy after $4.10^6$ links queries on the \contact network: the number $m'$ of discovered links; the percentage of tested pairs of nodes; the percentage of existing links found; and the efficiency coefficients $\mathcal{E}$ \mbox{and $\mathcal{R}$}.
}
\label{tab:eff-1000}
\end{table}

\subsection{Impact of the initial phase}

\begin{figure*}[ht]
\centering
\includegraphics[width=5.6cm]{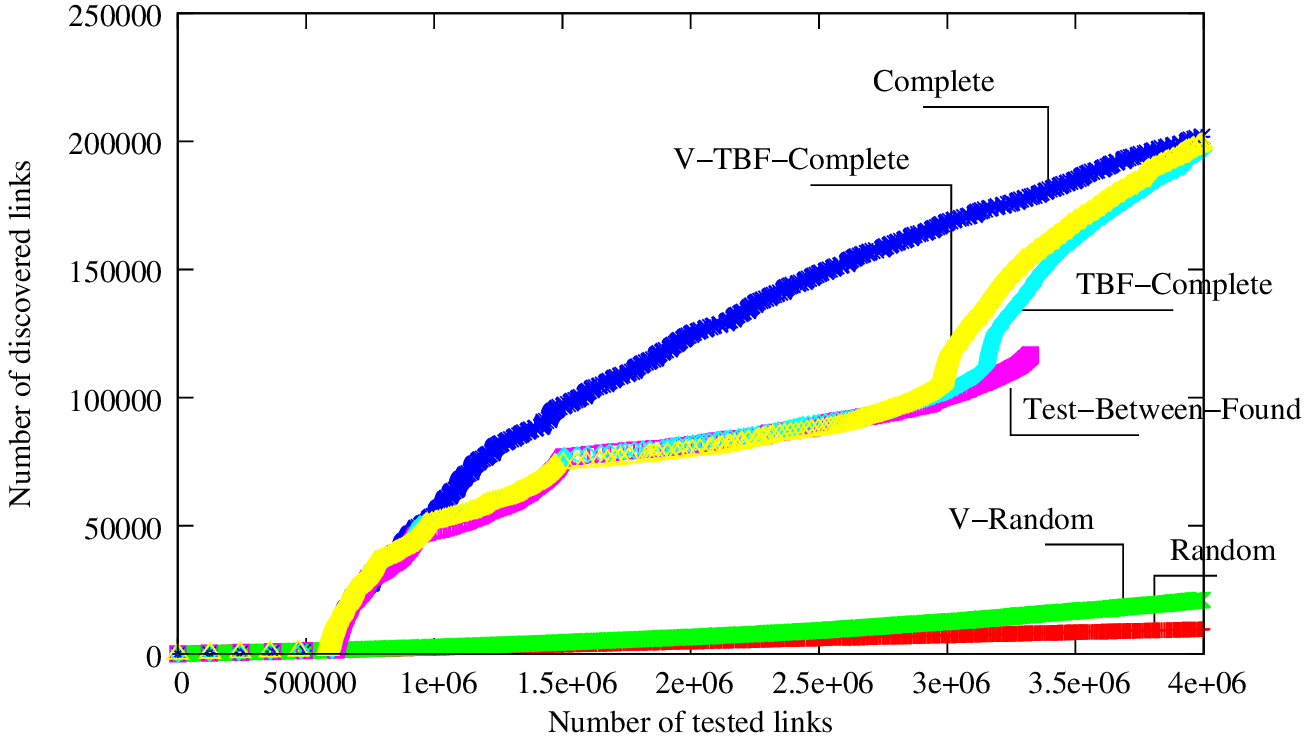}
\hspace{0.4cm}
\includegraphics[width=5.4cm]{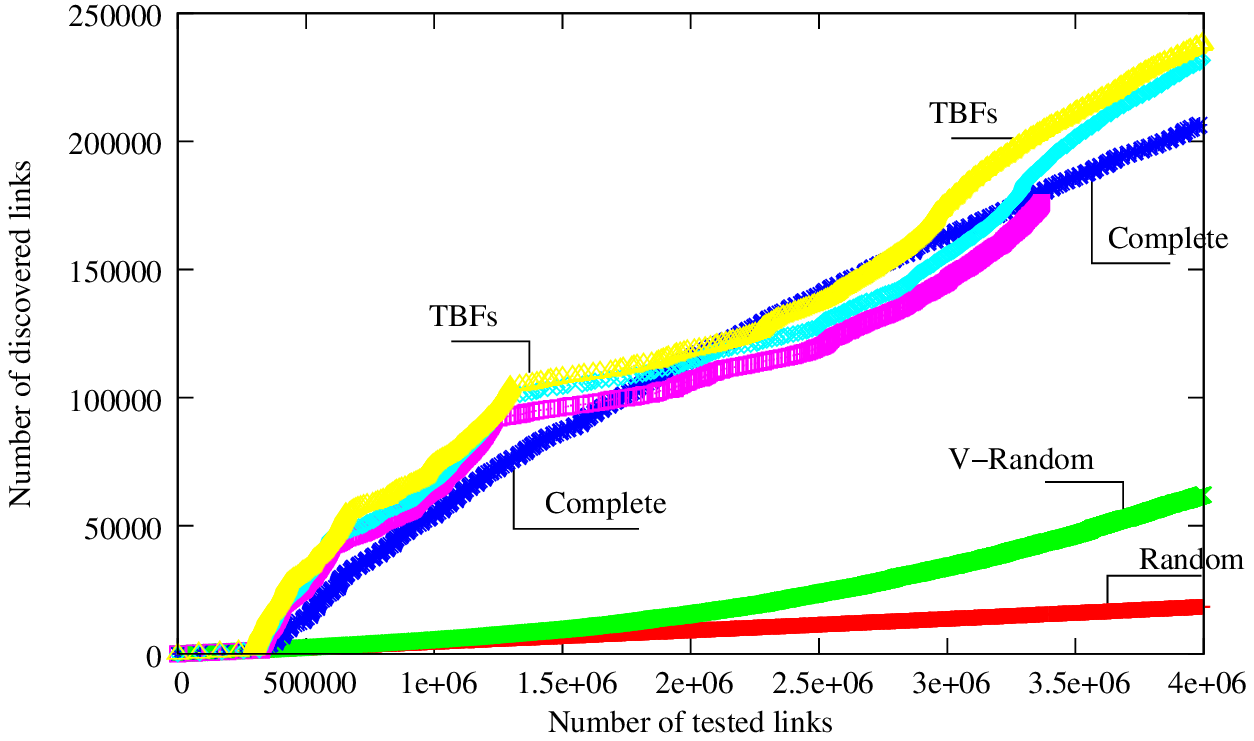}
\hspace{0.4cm}
\includegraphics[width=5.4cm]{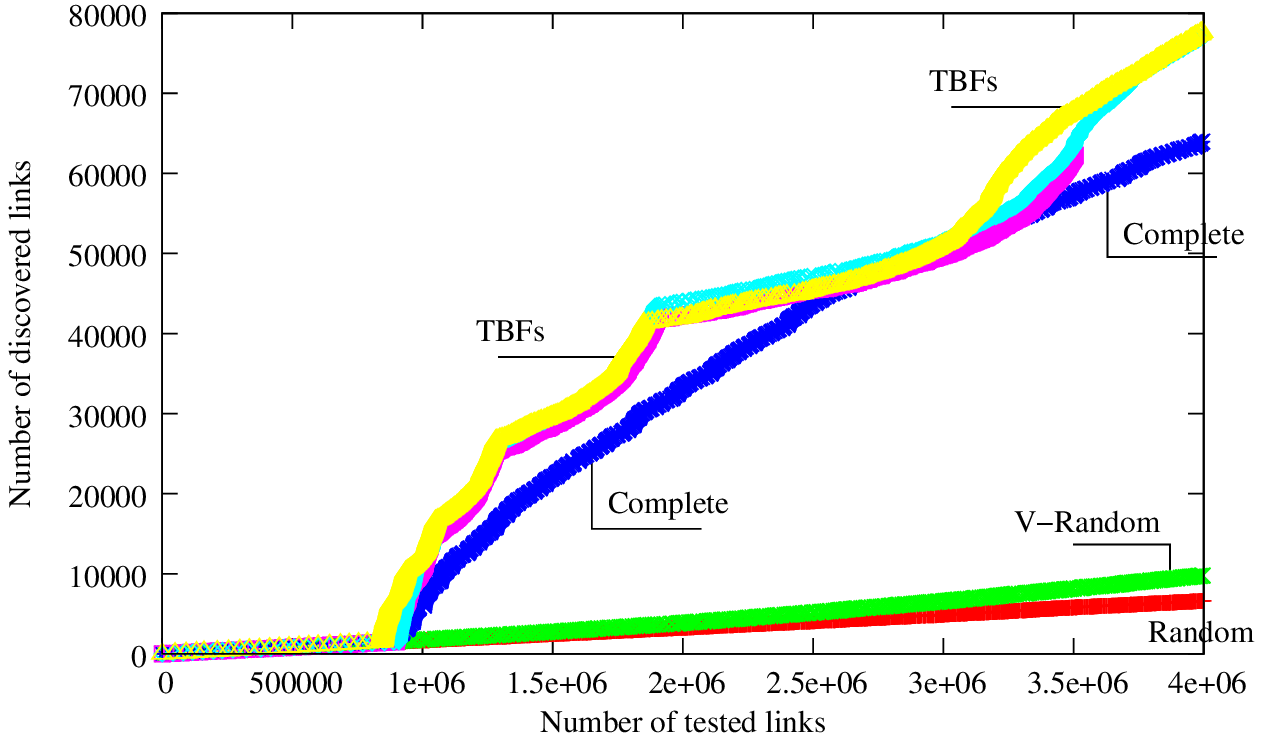}
\caption{
Number of links (vertical axis) discovered by each strategy as a function of the number of link queries performed (horizontal axis) for each of our three graphs (from left to right: \contact, \comment and \symmcomment), with $k=1\,500$. The \strat{tbf}{}, \strat{tbf-complete}{} and \strat{\V-tbf-complete}{} strategies are indicated as a unique curve (named TBFs) in the two last plots as the three curves overlap eachother.
}
\label{fig:curve-1500}
\end{figure*}

In order to test the impact of the initial phase on the efficiency of
the strategies, we conducted a similar experiment in which we
increased the parameter $k$ to $1\,500$. We present the results in
Figure~\ref{fig:curve-1500}. The change of the $k$ value implies in
particular that the random phase will last longer than in the previous
runs as it looks for a larger set of discovered links.  This induces a
delay before the beginning of the second phase of the strategies which
should in turn decrease their efficiency as they have less queries to
test the existence of the links.


Surprisingly though, the efficiency of the strategies does not seem to
be affected. The amount of discovered links after the same number of
queries is for instance comparable in the \contact network case
(around $21 \%$ of the existing links). This can be explained by the
fact that while searching for the $1\,500$ links, the random phase
has improved the partial knowledge of the network topology.
It is very likely then that the highly connected nodes have emerged
more significantly during this phase. Thus the ordering used by the
elaborated strategies, based on the \degreeprinciple, is in turn more
pertinent.


The plots based on the \comment and \symmcomment networks also show
that the behaviour of the strategies can be very similar in some
cases. This suggests to investigate other criteria to sort out their
efficiency. 

\subsection{Measurement bias}\label{sec:qual}

\begin{figure}[h!]
\begin{center}
\includegraphics[width=3.cm, angle=-90]{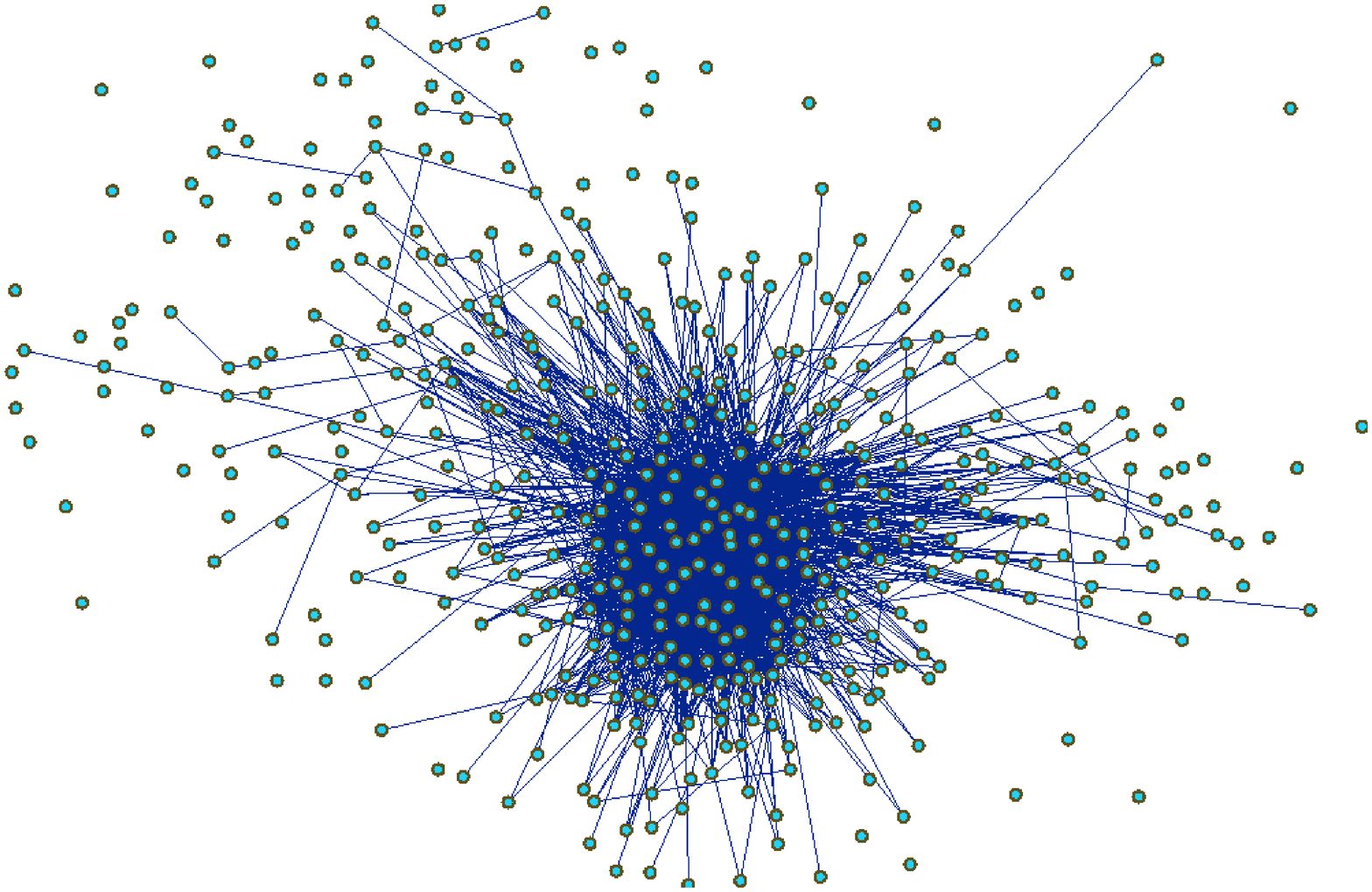}
\includegraphics[width=3.cm, angle=-90]{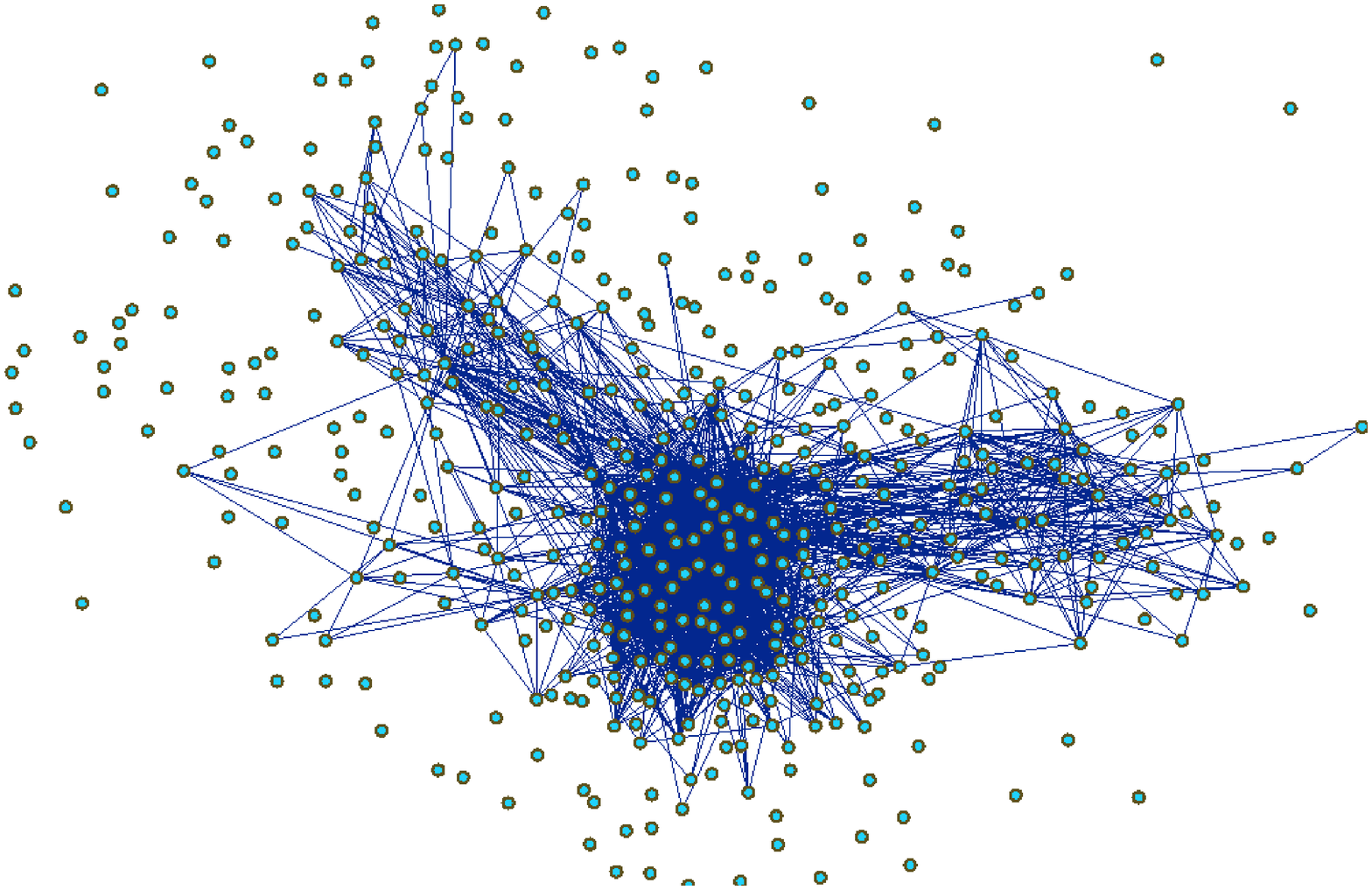}
\end{center}
\caption{
Drawings of samples obtained with the \strat{complete}{} (left) and \strat{tbf-complete}{} (right) strategies after the $20\,000$ link queries. The position of the nodes is the same in the two drawings (it is obtained by a classical graph drawing algorithm ran on the actual network), which makes it possible to observe visually that the links discovered by each strategy are not the same.
}
\label{fig:qual}
\end{figure}

Until now, we focused on our ability to discover many links with as few link queries as possible. However, different strategies discover different links, which may have consequences on the properties of the obtained samples: they may be biased by the measurement strategy, and biased differently depending on the strategy we use. This can be observed visually in Figure~\ref{fig:qual} for instance, and confirmed by the statistics given in Table~\ref{tab:qual}.

\begin{table}[h!]
\centering
\begin{tabular}{|c|c|c|c|c|c|c|}
  \hline
  & $m'$ & $\delta$ & avg deg & max deg & cc & tr\\
  \hline
  Reference& 21298 &0.002 &35.5 &1708 &0.083 &0.124 \\
  \hline
  \strat{random}{}& 6307 &0.000 &2.1 &38 &0.001 &0.001 \\
  \hline
  \strat{\V-random}{}& 6248 &0.001 &3.1 &123 &0.133 &0.120 \\
  \hline
  \strat{c}{1500}& 9840 &0.001 &13.0 &1708 &0.061 &0.422 \\
  \hline
  \strat{tbf}{1500} &2289 &0.024 &54.5 &663 &0.175 &0.208 \\
  \hline
  \strat{tbfc}{1500} &7717 & 0.003 & 20.0 & 1708& 0.085 &0.371 \\
  \hline
  \strat{\V-tbfc}{1500} & 8789& 0.002& 17.7& 1708 & 0.072 & 0.388 \\
  \hline
\end{tabular}
\caption{
\em \em
Main statistical properties (number of links finally discovered, density $\delta$, average degree, maximal degree, clustering coefficient and transitivity ratio) of the samples obtained by each measurement strategy with $k=1500$ applied on the \symmcomment network. We also display the properties of the actual network (first row), for comparison.
}
\label{tab:qual}
\end{table}

These experiments clearly show that the observed properties are biased
by the measurement (they are not the same as the ones of the actual
network), and moreover that different strategies lead to different
bias.

One can notice for instance that the \strat{complete}{} and the
\strat{test-between-found}{} strategies induce a very different bias
on the properties. It is likely to be due to the fact that the number
of involved nodes ($m'$) is very low for the second strategy but all
the possible links between them have been tested. This means in
particular that all the possible triangles have been discovered which
leads naturally to an over-evaluation of the clustering
coefficient. The strict opposite happens in the \strat{complete}{}
case since many nodes are discovered but the links between them are
not directly tested.

In some specific cases though, the values are correctly evaluated by
the strategies. Strategy \strat{\V-random}{}, for instance, gives a
correct value of the transitivity ratio. This is well explained by the
strategy itself that tests the existence of the third link of a
triangle as soon as two nodes appear to have a common
neighbor. 

It is also worth noticing that the mixed strategies have a better
evaluation of the clustering coefficient than other strategies. This
can be explained by the fact that, as the name suggests, they mix the
effects of the different strategies. In particular, the
over-evaluation of this property given by the
\strat{test-between-found}{} phase seems to be compensated by the
under-evaluation of the \strat{complete}{} phase.


These observations suggest to put in perspective the quantitative
assessments of the runs and to try integrating the qualitative point
of view in the evaluation of the efficiency of the strategies.





\section{Conclusion and perspectives}
\label{sec-conclu}

In this paper, we studied the problem of measuring large complex
networks when the measurement operation consists in testing the
existence of a link between two nodes. We proposed different
strategies for ordering the link queries in order to minimize their
number while maximizing the number of discovered links.
Those strategies rely on the expected statistical properties of the
network in order to predict the existence of the links and we tested
this approach on several real-world networks based on the Flickr
database.

The empirical results confirmed that the principles underlying the
development of the strategies are relevant in this measurement
context. The experiments showed that the elaborated strategies made a
huge improvement compared to the random approach. But they also raised
the question of accounting for the bias they induce on the extracted
samples. It turned out that the different strategies gave different
evaluations of the statistical properties of the original
networks. This result suggests to try combining them in order to
compensate those unwanted effects, which is what we plan to
investigate more specifically in the future.

Going further, we also intend to extend the kind of real-world
networks on which test the strategies, add measurement properties to
the list of statistical properties considered (such as the
assortativity and the degree-degree correlation) and try to adapt the
strategies to the directed graphs.  Another interesting perspective
would be to address the problem by means of formal approaches.






\bibliographystyle{IEEEtran}
\bibliography{biblio}

\end{document}